\documentclass[journal,twoside,web]{ieeecolor}
\usepackage{lcsys}

\usepackage{cite}
\usepackage{amsmath,amssymb,amsfonts}
\usepackage{graphicx}
\usepackage{textcomp}
\usepackage{xcolor}

\usepackage{amsthm}
\usepackage{algorithm}
\usepackage{algpseudocode}
\usepackage{tikz}
\usepackage{longtable}
\usepackage{multirow}
\usetikzlibrary{shapes,positioning}
\usepackage{etoolbox}

\newtheorem{lemma}{Lemma}
\newtheorem{remark}{Remark}
\AtEndEnvironment{problem}{\hfill$\Box$}
\AtEndEnvironment{lemma}{\hfill$\Box$}
\AtEndEnvironment{theorem}{\hfill$\Box$}
\AtEndEnvironment{assumption}{\hfill$\Box$}
\AtEndEnvironment{condition}{\hfill$\Box$}
\AtEndEnvironment{remark}{\hfill$\Box$}
\AtEndEnvironment{corollary}{\hfill$\Box$}
\AtEndEnvironment{stassumption}{\hfill$\Box$}
\AtEndEnvironment{definition}{\hfill$\Box$}
\newcommand{\vect}{\mathrm{vec}}

\newtheoremstyle{mystyle}
  {\topsep}
  {\topsep}
  {\itshape}
  {}
  {\bfseries}
  {.}
  { }
  {}

\title{\LARGE \bf Attitude Estimation Using Scalar Measurements}

\author{H. Alnahhal, S. Benahmed, S. Berkane, \textit{Senior Member, IEEE}, T. Hamel, \textit{Fellow, IEEE}
\thanks{Soulaimane Berkane is with the Department of Computer Science and Engineering, Université du Québec en Outaouais (UQO), QC J8X 3X7, Canada (soulaimane.berkane@uqo.ca).}
\thanks{Hassan Alnahhal is an independent researcher based in Egypt (h.alnahhal1989@gmail.com).}
\thanks{Sifeddine Benahmed is with the Department of Technology \& Innovation, Capgemini Engineering, Toulouse, 31300, France (sif-eddine.benahmed@capgemini.com).}
\thanks{Tarek Hamel is with I3S-UniCA-CNRS, Université Côte d’Azur, Sophia Antipolis, and the Institut Universitaire de France (thamel@i3s.unice.fr).}
\thanks{*This research is supported in part by NSERC-DG RGPIN-2020-04759 and the Fonds de recherche du Québec (FRQ).}}
\begin{document}
\maketitle
\thispagestyle{empty}
\pagestyle{empty}
\begin{abstract}
This paper revisits the problem of orientation estimation for rigid bodies through a novel framework based on scalar measurements. Unlike traditional vector-based methods, the proposed approach enables selective utilization of only the reliable axes of vector measurements while seamlessly incorporating alternative scalar modalities such as Pitot tubes, barometers with range sensors, and landmark-based constraints. The estimation problem is reformulated within a linear time-varying (LTV) framework, allowing the application of a deterministic linear Kalman filter. This design guarantees Global Uniform Exponential Stability (GES) under the Uniform Observability (UO) condition. Simulation results demonstrate the effectiveness of the proposed approach in achieving robust and accurate attitude estimation, even with partial vector measurements that simulate sensor axis failure.
\end{abstract}


\section{Introduction}
\IEEEPARstart{T}{he} estimation of a rigid body's orientation, or attitude, is a fundamental requirement in many engineering applications. It plays a critical role in diverse areas, including navigation, control of aerial and ground vehicles, and spacecraft stabilization. Attitude estimation is typically performed either statically, using body-frame measurements of known inertial vectors \cite{shuster1981three,markley1988attitude}, or dynamically, by fusing angular rate measurements with these vectors \cite{Mahony_Hamel_Pflimlin,batista2012sensor,zlotnik2016nonlinear}. Various sensors are employed to measure known inertial vectors in the body frame. For instance, accelerometers are used to detect the direction of gravity, providing a reference for pitch and roll angles, particularly effective when inertial accelerations are negligible compared to gravity. Magnetometers measure the Earth's magnetic field, offering a reliable heading reference when not affected by magnetic disturbances. These vector measurements, when combined with gyroscopes, enable robust dynamic attitude estimation--an approach that is prevalent in high-precision applications such as aerospace navigation, robotics, and unmanned aerial vehicles, where accurate orientation is essential
\cite{tayebi2006attitude,zlotnik2016nonlinear,Mahony_Hamel_Pflimlin,batista2012sensor,izadi2014rigid,crassidis2007survey}.

To achieve this sensor fusion, various attitude estimation approaches have been developed in the literature. Among these,  Kalman-type filters have been widely adopted in aerospace applications due to their ability to fuse sensor data effectively. However, they usually rely on linear approximations, which necessitate careful implementation \cite{crassidis2007survey}. Moreover, invariant Kalman filters \cite{barrau2017invariant,barrau2018invariant} have proven to be a robust solution, offering local asymptotic stability and addressing several limitations of traditional Kalman approaches.
Alternatively, recent research has focused on developing nonlinear deterministic observers that offer stronger stability guarantees and are better suited for handling the nonlinearities inherent in Inertial Navigation Systems (INS) applications, see for example \cite{hua2013implementation,zlotnik2016nonlinear,Mahony_Hamel_Pflimlin,berkane2017design,berkane2021nonlinear,Wang_TAC_2022}. These methods, though effective, largely depend on full three-dimensional vector information, which may not always be available or reliable in real-world conditions due to sensor limitations, environmental disturbances, or noise.

In this work, we propose a novel framework for rigid-body orientation estimation that leverages scalar measurements to enhance traditional methods. Unlike conventional vector-based approaches, scalar measurements impose partial constraints on the attitude without requiring full directional information. This allows the attitude estimator to remain operational even when vector measurements are unavailable or some axes are unreliable, addressing a critical gap in existing methods. The framework not only supports partial vector measurements but also enables the use of innovative sensor modalities to derive attitude constraints. For instance, altitude constraints can be obtained by two landmarks measurements of known height difference, or by fusing Z-range sensor data with barometric measurements for enhanced tilt estimation. 

To achieve this, we reformulate the estimation problem within a linear time-varying (LTV) framework that seamlessly accommodates both scalar and vector observations. While vector measurements naturally decompose into scalar components, our approach generalizes this concept to integrate diverse scalar measurements from various sensors. This reformulation enables the use of a deterministic linear Kalman filter.
The LTV framework, previously explored in attitude estimation contexts \cite{batista2012ges,batista2012globally,batista2012sensor,benahmed2024universal}, is further exploited here to guarantee global uniform exponential stability (GES) through rigorous Uniform Observability (UO) analysis.

\section{NOTATION}
We define \(\mathbb{Z}_{>0}\) as the set of positive integers, \(\mathbb{R}\) as the set of real numbers, \(\mathbb{R}^n\) as the \(n\)-dimensional Euclidean space, and \(\mathbb{S}^n\) as the \(n\)-dimensional unit sphere embedded in \(\mathbb{R}^{n+1}\). The Euclidean norm of a vector \( x \in \mathbb{R}^n \) is denoted by \(\|x\|\). For a vector \( x \in \mathbb{R}^n \), \( x_i \) represents its \(i\)-th element. We denote the \(n \times n\) identity matrix by \( I_n \) and the \(n \times n\) zero matrix by \( 0_{n \times n} \). The unit vectors along the coordinate axes are \( e_1 \), \( e_2 \), and \( e_3 \). The Special Orthogonal group of order three, denoted by \(\mathrm{SO}(3)\), is defined as \(\mathrm{SO}(3) := \{ A \in \mathbb{R}^{3 \times 3} : \det(A) = 1; \, AA^\top = A^\top A = I_3 \}\). The associated Lie algebra, denoted by \(\mathfrak{so}(3)\), is defined as \(\mathfrak{so}(3) := \{ \Omega \in \mathbb{R}^{3 \times 3} : \Omega = -\Omega^\top \}\). For vectors \( x, y \in \mathbb{R}^3 \), the mapping \([.]_\times : \mathbb{R}^3 \rightarrow \mathfrak{so}(3)\) is defined such that \([x]_\times y = x \times y\), where \(\times\) denotes the vector cross-product in \(\mathbb{R}^3\). The Kronecker product of two matrices \(A\) and \(B\) is represented by \( A \otimes B \). The vectorization operator \(\operatorname{vec} : \mathbb{R}^{m \times n} \rightarrow \mathbb{R}^{mn}\) stacks the columns of a matrix \( A \in \mathbb{R}^{m \times n} \) into a single column vector in \(\mathbb{R}^{mn}\). Conversely, the inverse vectorization operator \(\operatorname{vec}^{-1}_{m,n} : \mathbb{R}^{mn} \rightarrow \mathbb{R}^{m \times n}\) reshapes an \(mn \times 1\) vector back into an \(m \times n\) matrix. Given a rotation vector $x\in\mathbb{R}^3\setminus\{0\}$, the
corresponding rotation matrix is given by the exponential map $\exp([x]_{\times})\in \mathrm{SO}(3)$ which
is given by the following 
Rodrigues formula:
\begin{equation}\label{eq:exp}
    \exp([x]_{\times})=I_3+\frac{\sin(\|x\|)}{\|x\|}[x]_{\times}+\frac{1-\cos(\|x\|)}{\|x\|^2}[x]_{\times}^2.
\end{equation}

\section{PROBLEM FORMULATION}
\begin{figure}[h!]    
\centering
    \includegraphics[width=0.5\textwidth]{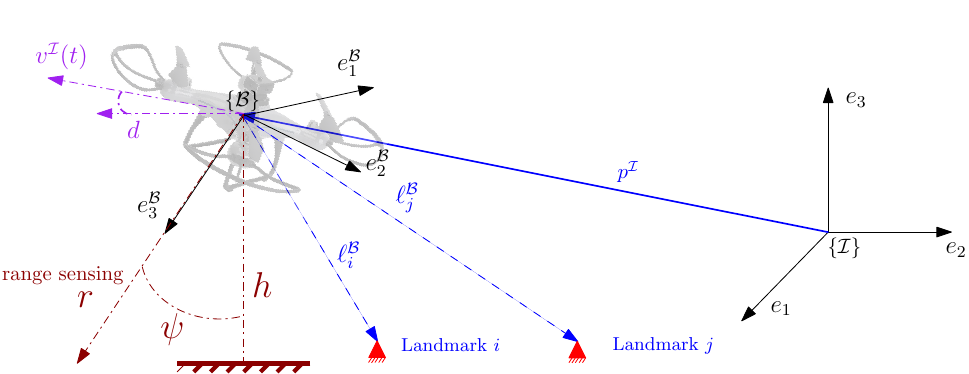}
\caption{Illustration of different sources of scalar measurements.}
\label{fig:Schematic_pitot_angle}
\end{figure}

Let \(\{ \mathcal{I} \}\) represent an inertial reference frame, and \(\{ \mathcal{B} \}\) be a body-fixed frame attached to the center of mass of a rigid body (\textit{e.g.,} a vehicle). The rotation matrix \( R \in \mathrm{SO}(3) \) defines the orientation (attitude) of frame \(\{ \mathcal{B} \}\) relative to \(\{ \mathcal{I} \}\). The 3D kinematic equation governing the attitude of a rigid body is given by:
\begin{align}
\dot{R} = R[\omega]_\times,
\label{orintation}
\end{align}
where \(\omega\) denotes the angular velocity of \(\{ \mathcal{B} \}\) with respect to \(\{ \mathcal{I} \}\), expressed in frame \(\{ \mathcal{B} \}\). In this work, we assume that the vehicle is equipped with a sensor providing measurements of the angular velocity $\omega$ (\textit{e.g.,} gyroscope). Furthermore, it is assumed available  a set of $q\geq 1$ \textit{scalar measurements} modelled as follows: 
\begin{align}\label{general_measurement_outputs}
y_i:= a^\top_i R^\top b_i,\quad i=1,\cdots, q, 
\end{align}
where $y_i$ denotes a general scalar attitude measurement, with $a_i, b_i \in \mathbb{R}^3$ representing known (possibly time-varying) vectors. A variety of sensors are capable of producing measurements in the form described by \eqref{general_measurement_outputs}. In the sequel, we outline several application examples that illustrate how such scalar attitude measurements can be derived, see Fig.~\ref{fig:Schematic_pitot_angle}:

\subsubsection{Complete or partial vector measurements}
Traditional attitude estimation is based on combining a set of body-frame measurements of known inertial directions (e.g., magnetometer, sun sensor, etc.). Assume the vehicle is equipped with sensors that provide body-frame components \( r^{\mathcal{B}} \in \mathbb{R}^3 \) of a known inertial vector \( r^{\mathcal{I}} \in \mathbb{R}^3 \), such that:
\begin{align}
    r^{\mathcal{B}} = R^\top r^{\mathcal{I}}.
    \label{Vector_ measurements}
\end{align}
Equation \eqref{Vector_ measurements} provides three scalar measurements in the form of \eqref{general_measurement_outputs}. Specifically, setting \( r^{\mathcal{B}} = [y_1, y_2, y_3]^\top \), we identify \( a_i = e_i \) and \( b_i = r^{\mathcal{I}} \) for \( i \in \{1,2,3\} \).

Importantly, the proposed framework also accommodates partial vector measurements, where only a subset of the components of \( r^{\mathcal{B}} \) is available. This is particularly useful in scenarios where certain sensor axes degrade or become temporarily unavailable. For instance, if only the first and third components of \( r^{\mathcal{B}} \) are accessible, the corresponding constraints with \( a_1 = e_1 \) and \( a_3 = e_3 \) can still be exploited independently in the estimation process.

\subsubsection{Tilt angle measurements}
Tilt angle measurements provide another source of scalar data for attitude estimation, particularly in aerial vehicles. 
By combining a barometer, which provides the vehicle's height $h$, with a down-facing range sensor that measures the range $r$ to the ground (assumed flat), the tilt angle $\psi$ can be determined as:
\begin{equation}\cos(\psi) = \frac{h}{r} = e_3^\top R^\top e_3,\end{equation}
which fits the measurement model \eqref{general_measurement_outputs}, where $a_i = b_i = e_3$.
\subsubsection{Landmark-based virtual measurements}

Attitude estimation can be enhanced using landmark measurements by exploiting the relative altitude difference between two known landmarks. Let \( p^{\mathcal{I}} \) denote the position of the vehicle in the inertial frame, and \( p^{\mathcal{I}}_{\ell_i} \) and \( p^{\mathcal{I}}_{\ell_j} \) represent the positions of two landmarks in \( \{\mathcal{I}\} \) (not necessarily known). The relative measurement in the body frame is given by:
\begin{align}
   \ell_i^{\mathcal{B}}(t) - \ell_j^{\mathcal{B}}(t) = R^\top (p^{\mathcal{I}}_{\ell_i} - p^{\mathcal{I}}_{\ell_j}).
    \label{relative_landmark_constraint}
\end{align}
By projecting this difference onto the vertical axis, we obtain:
\begin{align}
   {(\ell_i^{\mathcal{B}}(t) - \ell_j^{\mathcal{B}}(t))^\top} R^\top e_3 = e_3^\top (p^{\mathcal{I}}_{\ell_i} - p^{\mathcal{I}}_{\ell_j}).
   \label{vertical_projection}
\end{align}
This relation serves as a virtual measurement that enforces alignment in the vertical axis. Importantly, this approach does not require absolute altitude measurements—only the assumption that the landmarks have a known height difference (\textit{e.g.,} if both are at the same elevation).


\subsubsection{Pitot tube measurements}
Pitot tube measurements, commonly used in aviation, provide scalar data related to the vehicle's airspeed and contribute to attitude estimation, especially for pitch angle determination \cite{Ariante_Ponte_Papa_DelCore,de2024pitot}. The Pitot tube measures the component of the vehicle's velocity in the direction of the probe, expressed in the \textit{body frame}. The scalar output of the Pitot tube is given by:
\begin{align}\label{pitot_angle}
    {y_p} := d^\top R^\top v^{\mathcal{I}}(t),
\end{align}
where $d$ is the known direction of the Pitot probe in the body frame and $v^{\mathcal{I}}(t)$ is the vehicle's velocity in the inertial frame. We assume that $v^{\mathcal{I}}(t)$ is obtained from GPS measurements under the condition of \textit{negligible wind}. This simplifies the relationship between the ground velocity (given by GPS) and the airspeed measured by the Pitot tube. When the wind speed is insignificant, the inertial velocity $v^{\mathcal{I}}(t)$ closely approximates the airspeed vector.  In our model, we express the measurement in the form of \eqref{general_measurement_outputs} by setting $a_i = d$ and $b_i = v^{\mathcal{I}}(t)$. This enables the use of the Pitot tube as an additional constraint in the attitude estimation process.

%

Overall, scalar measurements, as formulated in \eqref{general_measurement_outputs}, provide a versatile and flexible framework for attitude estimation. By not limiting the estimation process to full vector measurements, this approach allows for the incorporation of data from a wide range of sensors, including those incapable of providing complete directional information. This adaptability enables the integration of innovative sensor combinations, enhancing the robustness and accuracy of the estimation process across various applications.

\section{MAIN RESULTS}
In this section, we formulate the attitude kinematics as a state-space model, where the states correspond to the columns of the transposed rotation matrix expressed in the body frame. The output vector is derived from the generic measurement model \eqref{general_measurement_outputs}, resulting in a linear time-varying (LTV) system. We then analyze the system's uniform observability (UO) and propose a Kalman-type observer that guarantees global exponential stability of the filter under the UO condition. To formalize this, the rotation matrix $R$ is represented by stacking the columns of $R^\top$ into the vector:
\begin{align}
 x^{\mathcal{B}}  := \operatorname{vec}(R^\top), 
 \label{Z}
\end{align}
where $\operatorname{vec}(\cdot)$ denotes the vectorization operator. The dynamics of this state vector, derived from \eqref{orintation}, are given by:
\begin{align}
\dot{x}^{\mathcal{B}} = - (I_3 \otimes [\omega]_\times) x^{\mathcal{B}}=:A(t)x^{\mathcal{B}}.
\label{X_dot}
\end{align}
By applying the identity $\operatorname{vec}(ABC) = (C^\top \otimes A) \operatorname{vec}(B)$, the general measurement model \eqref{general_measurement_outputs}  can be reformulated into a linear output equation:
\begin{align}
y_i &= (b_i^\top \otimes a_i^\top )x^{\mathcal{B}} .
\label{general_output}
\end{align}
We can now define the following output vector:
\begin{equation}\label{equation:output_vector}    y=\begin{bmatrix}
        y_1\\
        \vdots\\
        y_q
    \end{bmatrix}=\begin{bmatrix}
        b_1^\top \otimes a_1^\top \\
        \vdots\\
        b_q^\top \otimes a_q^\top 
    \end{bmatrix}x^{\mathcal{B}}=:C(t)x^{\mathcal{B}}.
\end{equation}
This formulation ensures the system is linear in the state 
$x^{\mathcal{B}}$, facilitating the design of the observer. 
In fact, in view of \eqref{X_dot} and \eqref{equation:output_vector}, we obtain the following LTV system
\begin{align}\label{equation:LTV_state_space}
    \dot{x}^{\mathcal{B}} = A(t)x^{\mathcal{B}}, \quad y = C(t)x^{\mathcal{B}}.
\end{align}
It is important to note that the matrix $A(t)$ is time-varying because it depends on the profile of the angular velocity $\omega(t)$, which acts as an external time-varying signal. Similarly, the matrix $C(t)$  is possibly time-varying, as it relies on potentially time-varying known vectors $a_i$ and $b_i$.


The state of system \eqref{equation:LTV_state_space} can then be estimated using a standard deterministic linear Kalman filter defined as:
\begin{align}
  \dot{\hat{x}}^\mathcal{B} = A(t)\hat{x}^\mathcal{B} + K(t)(y - C(t)\hat{x}^\mathcal{B}), 
  \label{recatii_obs}
\end{align}
where $\hat{x}^\mathcal{B}$ denotes the estimate of $x^\mathcal{B}$, and the gain matrix $K(t)$ is given by $K(t) = P C(t)^\top Q(t), 
$
with $P$ is the solution to the continuous Riccati equation (CRE):
\begin{align*}
\dot{P} = A(t)P + P A^\top(t) - P C^\top(t)
Q(t) C(t) P + M(t),
\end{align*}
where \( P(0) \) is a positive definite matrix, and \( Q(t) \) and \( M(t) \) are uniformly positive definite matrices to be specified (see also Remark \ref{remark_constant_gain}). The stability and convergence properties of this filter are related to the well-posedness of the solution \( P \) of the CRE, which is itself connected to the UO property. In fact, as discussed in \cite{Hamel_Samson}, if we define the estimation error 
$
\tilde{x}^{\mathcal{B}} := x^{\mathcal{B}} - \hat{x}^{\mathcal{B}}
$
and consider the Lyapunov function 
$
V = (\tilde{x}^{\mathcal{B}})^\top P^{-1} \tilde{x}^{\mathcal{B}},
$
then we have:
\[
\dot{V}= -(\tilde{x}^{\mathcal{B}})^\top \left(C^\top(t)
Q(t) C(t)+P^{-1} M P^{-1}\right) \tilde{x}^{\mathcal{B}},
\]
which is negative definite if both \( P \) and \( P^{-1} \) are well-defined and positive definite.

\begin{remark}[\textbf{Fixed-gain design}] \label{remark_constant_gain}
    Since \( A(t) \) is a skew-symmetric matrix, it is straightforward to verify that \( P = I_9 \) is a valid solution to the CRE \eqref{recatii_obs} by choosing \( M = C^\top Q C \), which is only positive semi-definite. In this case, the observer gain becomes \( K = C^\top Q \), and we have $\dot{V}= -2(\tilde{x}^{\mathcal{B}})^\top C^\top Q C \tilde{x}^{\mathcal{B}}$, which can be shown to imply UGES under the same UO condition; see also \cite{loria2002uniform}.
\end{remark}

\begin{remark}[\textbf{Tuning of $M$ and $Q$}] \label{remark_tunning_Q_V}
    The proposed observer is deterministic, but the matrices \( M(t) \) and \( Q(t) \) can be locally tuned based on the noise properties of the measurements, enabling an estimation performance that is optimal or near-optimal in the spirit of the stochastic Kalman filter. Assume the gyroscope measures \( \omega^y = \omega + n^{\omega} \), where \( n^{\omega} \) is additive noise. Substituting into \eqref{equation:LTV_state_space} yields:
    \[
    \dot{x}^{\mathcal{B}} = A^y(t)x^{\mathcal{B}} + \mathcal{N}(\hat{x}^{\mathcal{B}}) n^{\omega}, \quad A^y(t) = - (I_3 \otimes [\omega^y]_\times),
    \]
    where \( \mathcal{N}(\hat{x}^{\mathcal{B}}) = -\begin{bmatrix}
        [e_1^{\mathcal{B}}]_{\times} & [e_2^{\mathcal{B}}]_{\times} & [e_3^{\mathcal{B}}]_{\times}
    \end{bmatrix}^{\top} \). Noise \( n^y \) in measurements modifies the output equation as \( y = Cx^{\mathcal{B}} + n^y \). By approximating \( \mathcal{N}(x^{\mathcal{B}}) \) with \( \mathcal{N}(\hat{x}^{\mathcal{B}}) \), the tuning becomes:
    \[
    M = \mathcal{N}(\hat{x}^{\mathcal{B}}) \mathrm{Cov}(n^{\omega}) \mathcal{N}(\hat{x}^{\mathcal{B}})^{\top}, \quad Q^{-1} = \mathrm{Cov}(n^y).
    \]
    To ensure uniform positive definiteness, a small positive definite term can be added to \( M \) and \( Q^{-1} \) if needed.
\end{remark}

Next, the orientation matrix is computed as: 
\begin{equation}
     \bar{R} = \left(\vect_{3,3}^{-1}(\hat{x}^{\mathcal{B}})\right)^\top.   
\end{equation}
Note that \(\bar{R}\) is not necessarily a valid rotation matrix, although it converges to one over time. To obtain a valid rotation matrix for all times, \(\bar{R}\) is projected onto the closest rotation matrix using Singular Value Decomposition (SVD), as detailed in \cite{sarabandi2023solution}. The SVD decomposition is given by:
\begin{equation}\label{equation:SVD_decomposition}
\bar{R} = U \Sigma V^\top,
\end{equation}
and the nearest rotation matrix to \(\bar{R}\) is:
\begin{equation}
\hat{R} = U D V^\top, \quad \text{with} \quad D = \mathrm{diag}(1, 1, \det(U V^\top)).
\end{equation}
Furthermore, to ensure consistency, the filter's state is updated at each time step with the nearest valid rotation matrix:
\begin{equation}\label{eq:reset}
    (\hat{x}^{\mathcal{B}})^+\leftarrow \text{vec}(\hat{R}^\top).
\end{equation}
The aforementioned projection and reset steps preserve the geometric structure of the orientation estimates and help improving the estimator's convergence (at least locally). In fact, since by definition $\hat{R} = \arg\min_{Q \in SO(3)} \|Q - \bar{R}\|_F$, the optimality condition imposes that the matrix $\hat R - \bar{R}$ is orthogonal to any tangent vector at $\hat R$ on the manifold $SO(3)$. 
Now, in a local neighborhood of the true rotation $R$, the matrix $R - \hat R$ lies approximately in the tangent space of $SO(3)$ at $\hat R$ and thus orthogonal to $\hat R - \bar{R}$. This orthogonality allows us to apply the Pythagorean theorem, yielding:
\[
\|R - \bar{R}\|_F^2 = \|R - \hat R\|_F^2 + \|\hat R - \bar{R}\|_F^2\geq\|R - \hat R\|_F^2.
\]
Therefore, the (optional) reset step in \eqref{eq:reset} strictly reduces or maintains the distance to the true rotation, ensuring (locally) an improved geometric consistency and estimator convergence.

\begin{figure}[h!]
    \centering
\includegraphics[width=\linewidth]{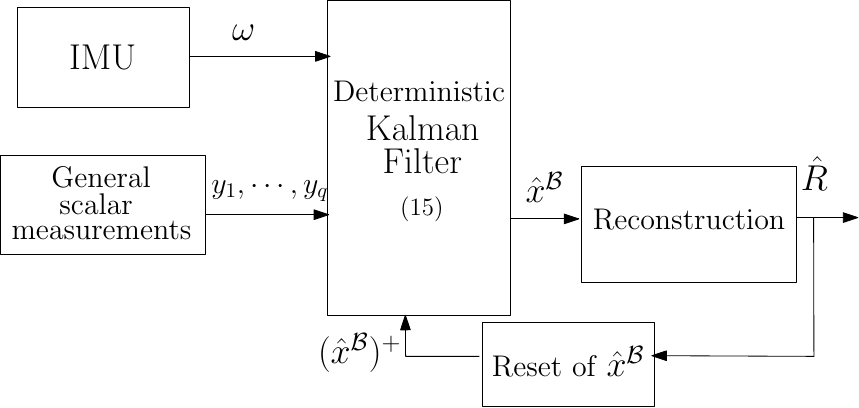} 
    \caption{Illustration of the proposed attitude estimation approach.}
    \label{fig:Proposed state estimation approach}
\end{figure}

Fig.~\ref{fig:Proposed state estimation approach} illustrates the structure of the proposed estimation approach. Based on the available measurements, the observer computes the estimates of $x^{\mathcal{B}}$. These estimates are then used to reconstruct the attitude $\hat{R}$ through SVD. Subsequently, $\hat{x}^{\mathcal{B}}$ is reset using reconstructed $\hat{R}$.  Next, we introduce a technical lemma which simplifies the analysis of uniform observability for the pair $(A(\cdot),C(\cdot))$ in the sense of \cite[Theorem 2.3]{Hamel_Samson}. Note that the uniform observability property guarantees uniform global exponential stability of the deterministic Kalman filter, see \cite{Wang_TAC_2022,Hamel_Samson,barrau2017invariant} for more details.
\begin{lemma}\label{Lemma:general_condition}
The pair \( (A(\cdot), C(\cdot)) \) is uniformly observable if and only if there exist $\delta,\mu>0$ such that: for any $t\geq0$,
\begin{align}\label{equation:condition_of_lemma_1}
\int_{t}^{t+\delta}\sum_{i=1}^qu_i(s)u_i(s)^\top ds\geq \mu I_9,
\end{align}
where $u_i(s):=(b_i(s)\otimes R(s)a_i(s))\in\mathbb{R}^9$.
\end{lemma}
\renewcommand\qedsymbol{$\blacksquare$}
\begin{proof}
See Appendix A.
\end{proof}
Lemma~\ref{Lemma:general_condition} establishes a uniform observability condition for the general measurement model presented in \eqref{general_measurement_outputs}. This equivalence result serves as a foundational tool for analyzing observability in various case studies, including vector and scalar measurements. For instance, the next lemma provides a sufficient condition in the case of vector measurements. 

\begin{lemma}\label{Lemma:vector_measurements}
Suppose available only $m\geq 1$ vector measurements $r^{\mathcal{B}}_j= R^\top r^{\mathcal{I}}_j$. Let $y$ in \eqref{equation:output_vector} where $[y_{3j-2}, y_{3j-1}, y_{3j}]^\top=r_j^{\mathcal{B}},$ $b_{3j-2}=b_{3j-1}=b_{3j}=r_j^{\mathcal{I}}$ and $a_{i+3(j-1)}=e_i$ with $i\in\{1,2,3\}$ and $j\in\{1,\cdots,m\}$. 
Then, the pair \( (A(\cdot), C(\cdot)) \) is uniformly observable if and only if there exist $\delta,\mu>0$ such that 
\begin{equation}\label{equation:condition_of_lemma_2}
\int_{t}^{t+\delta}\sum_{j=1}^m r_j^{\mathcal{I}}(s) (r_j^{\mathcal{I}}(s))^\top ds\geq \mu I_3,    
\end{equation}
for any $t\geq0$.
\end{lemma}
\begin{proof}
See Appendix B.
\end{proof}
The condition in Lemma~\ref{Lemma:vector_measurements} is satisfied if the set of measurements includes at least three constant non-collinear inertial vectors. This condition can also be met with only two non-collinear inertial vectors. Indeed, given two non-collinear vectors $ r_1^{\mathcal{I}}$ and $r_2^{\mathcal{I}}$, with their respective measurements $ r_1^{\mathcal{B}}$ and $ r_2^{\mathcal{B}}$, one can construct a third vector $ r_3^{\mathcal{I}} = r_1^{\mathcal{I}} \times r_2^{\mathcal{I}} $ using their cross product. This ensures the existence of a third non-collinear inertial vector with the corresponding measurement $r_3^{\mathcal{B}} = r_1^{\mathcal{B}} \times r_2^{\mathcal{B}}$, thus satisfying the condition. In the case of time-varying inertial vectors, the condition can still be satisfied with a single vector, provided that it exhibits persistent excitation (PE). These results are consistent with established observability criteria; see \cite{Mahony_Hamel_Pflimlin,trumpf2012analysis}.


 \section{Discretization and Implementation Aspects}\label{section:discreet}
In this section, we present a closed-form discrete version of the proposed observer. To derive the discrete propagation of the observer, we leverage the matrix exponential of the continuous-time state matrix
$A(t)$ defined in \eqref{X_dot}. Under the Zero-Order Hold (ZOH) assumption, which considers \( \omega(t) \) constant over the sampling interval, the discrete-time state transition matrix becomes:
\[
A_k := \exp(A(k\tau)\tau) = I_3 \otimes \exp(-[\omega_k \tau]_\times).
\]
This expression provides a closed-form discrete representation of the continuous dynamics, leveraging the matrix exponential of the skew-symmetric matrix given in 
 \eqref{eq:exp}. Algorithm~\ref{Algorithm_1} presents the implementation of the closed-form discrete version of the proposed observer, structured as a correction-prediction scheme. It accounts for sensor data arriving at different frequencies, particularly the higher update rate of the IMU compared to other sensors. This discrepancy arises due to differences in hardware design and processing requirements.

Following the prediction-correction framework, the IMU is responsible for the prediction step, while other sensors contribute to the correction step. Accordingly, the discrete version of matrix $M$ is tuned based on the IMU update rate $f_{\mathrm{IMU}}$, ensuring accurate state propagation, while the discrete version of matrix $Q$ is adjusted according to the update rate of each respective sensor $f_{\mathrm{sensor}}$. The resulting discrete formulations are: $
    M_k = (1 / f_{\mathrm{IMU}}) M(k\tau), \quad
    Q_k^{-1} = (1 / f_{\mathrm{sensor}}) Q^{-1}(k\tau),$ where $M(k\tau)$ and $Q^{-1}(k\tau)$ are tuned as described in Remark~\ref{remark_tunning_Q_V}.

\begin{algorithm}[h!]
\caption{Discrete-time implementation of the proposed deterministic Kalman filter for attitude estimation}
 \label{Algorithm_1}
 \begin{algorithmic}[1]
\Statex \textbf{Input:}  $\hat{x}^{\mathcal{B}}_{0|0}$, $P_{0|0}$, $\tau$, $y_k$ 
\Statex \textbf{Output:} $\hat{R}_k$, for any $k\in\mathbb{Z}_{\geq1}$.\vspace{1mm} 
\hrule height 1pt 
\vspace{1mm} 
\State $\bar{R}_0 \leftarrow (\mathrm{vec}^{-1}_{3,3}(\hat{x}^{\mathcal{B}}_{0|0}))^{\top}$ 
\For{each time step $k\geq1$} \hfill
\Statex \texttt{\textcolor{blue}{/$\star$ \textbf{Prediction Step:} $\star$/}}\hfill
 \If{IMU data $\omega_{k-1}$ is available}
    \State   $A_{k-1}\leftarrow I_3\otimes \exp(-[\omega_{k-1}\tau]_\times)$
   \State    $\hat{x}_{k|k-1}^{\mathcal{B}} \leftarrow A_{k-1} \hat{x}_{k-1|k-1}^{\mathcal{B}}$
  \State  $P_{k|k-1} \leftarrow A_{k-1} P_{k-1|k-1} A_{k-1}^\top +M_k$ \hfill 
    \EndIf 
\Statex \texttt{\textcolor{blue}{/$\star$ \textbf{Update Step:} $\star$/}}
   \If {Sensor data is available}
       \State Compute the matrix $C_k$ \texttt{\textcolor{blue}{/$\star$ use \eqref{equation:output_vector} $\star$/}}
     \State $K\leftarrow P_{k|k-1}C^{\top}_k(C_kP_{k|k-1}C^{\top}_k+Q_k^{-1})^{-1}$
     \State $\hat{x}_{k|k}^{\mathcal{B}} \leftarrow \hat{x}_{k|k-1}^{\mathcal{B}}+K(y_k-C_k\hat{x}_{k|k-1}^{\mathcal{B}})$
     \State $P_{k|k}\leftarrow(I_{9}-KC_k)P_{k|k-1}$
      \Else
    \State  $\hat{x}_{k|k}^{\mathcal{B}} \leftarrow \hat{x}_{k|k-1}^{\mathcal{B}}$
      \State  $P_{k|k}\leftarrow P_{k|k-1}$
        \EndIf   
   \State $P_{k|k}\leftarrow\frac{1}{2}(P_{k|k} + P_{k|k}^{\top})$

  \Statex \texttt{\textcolor{blue}{/$\star$ \textbf{Attitude Reconstruction:} $\star$/}} 
  \State $\bar{R}_k \leftarrow (\mathrm{vec}^{-1}_{3,3}(\hat{x}_{k|k}^{\mathcal{B}}))^{\top}$ 
   \State Compute $U_k$ and $V_k$ using SVD decomposition of $\bar{R}_k$ as in \eqref{equation:SVD_decomposition}.
   \State $D_k\leftarrow \textrm{diag}(1,1,\det(U_kV_k^T))$
    \State    $\hat{R}_k\leftarrow U_kD_kV_k^{T}$
   \Statex \texttt{\textcolor{blue}{/$\star$\textbf{ Reinitialize the value of $\hat{x}_{k|k}^{\mathcal{B}}$} $\star$/}}
   \State $\hat{x}_{k|k}^{\mathcal{B}}=\mathrm{vec}(\hat{R}_k^{\top})$
  \EndFor
\end{algorithmic}
\end{algorithm}
\section{Simulation Results}
In this section, we provide simulation results to test the performance of the proposed observer.
Consider a vehicle moving in 3D space. 
The rotational motion of the vehicle is subject to the following angular velocity 
$\omega(t)=\begin{bmatrix}\sin(0.3t) &0.7\sin(0.2t+\pi) &0.5\sin(0.1t+\pi/3)\end{bmatrix}$ and the initial value of the true attitude is $R(0)=\exp([\pi e_2]_{\times}/2)$. We consider that the vehicle is equipped  with a three-axis gyroscope, a three-axis accelerometer and a three-axis magnetometer. The magnetometer reading in the body frame are given by $m^{\mathcal{B}}=R^{\top}[\tfrac{1}{\sqrt{2}},0,\tfrac{1}{\sqrt{2}}]^{\top}$. Under the common assumption of negligible translational accelerations, the accelerometer measurements in the body frame is given by $a^{\mathcal{B}}=-R^{\top}ge_3$ with $g=9.81$[m/s$^2$]. 

We examine three measurement‐availability scenarios to test robustness under partial information. As detailed in Table~\ref{table:cases_simulation}, in each case, the observer receives angular-velocity data and full or partial body‐frame measurements of accelerometer  and magnetometer. To reflect the discrepancies in sensor frequencies, the generated measurements data is downsampled at different frequencies, see Table~\ref{table:cases_simulation}. The discrete-time version of the observer is implemented as in Algorithm~\ref{Algorithm_1}. The matrices $M$ and $Q$ are tuned as in Section~\ref{section:discreet}, along with the sensor covariances specified also specified in Table~\ref{table:cases_simulation}.

\begin{figure}[h!]
    \centering
    \includegraphics[width=\linewidth]{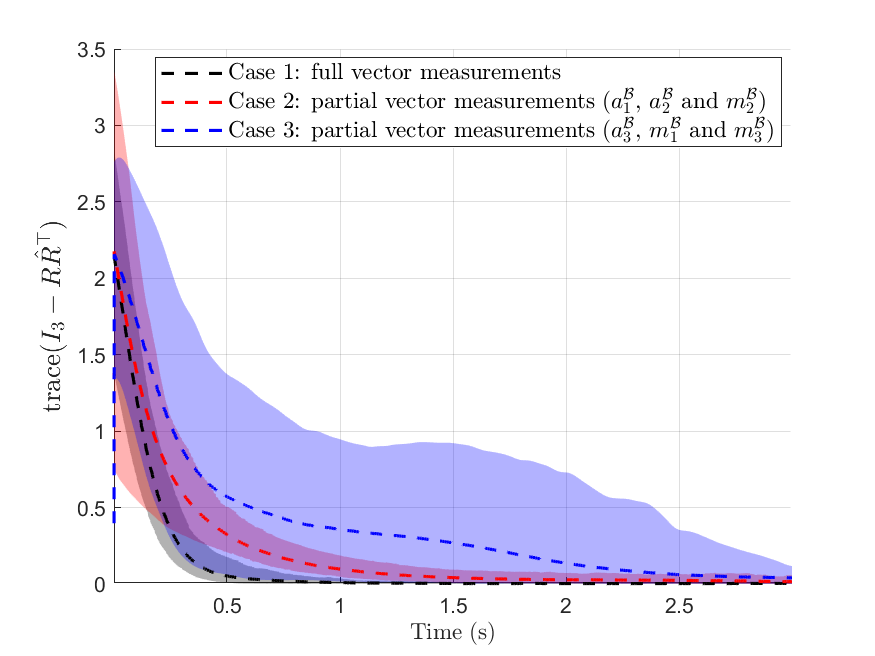} 
    \caption{Average attitude estimation errors.}
    \label{Figure:simulation_results}
\end{figure}

We conduct a Monte-Carlo simulation with $100$ runs. For each run, the initial attitude estimation errors are randomly generated using Gaussian distributions with an average attitude error corresponds to $22.5$(deg) errors in roll, pitch and yaw. The average estimation errors are presented in Fig.~\ref{Figure:simulation_results}. The shaded areas illustrate
the $5$th to $95$th percentile of error. Despite the reduced information in Cases 2 and 3, the observer achieves exponential convergence in all trials, confirming its ability to reconstruct full attitude using partial vector measurements. Therefore, the proposed approach offers enhanced reliability in  environments where one or two sensor axis are compromised whether by hardware failure, magnetic disturbance or similar anomalies. 

Finally, note that the percentile band for Case~3 is substantially wider than that for Case~2. This larger dispersion reflects the heavier reliance of Case 3 on magnetometer readings, which exhibit both higher noise variance and lower sampling frequency compared to accelerometer data.

\begin{table}[h!]
\centering
\caption{Available measurements considered for each case.}
\begin{tabular}{|c|c|c|c|c|c|c|c|}
\hline & \textbf{$\omega$}
 & \textbf{$a^{\mathcal{B}}_1$} & \textbf{$a^{\mathcal{B}}_2$} & \textbf{$a^{\mathcal{B}}_3$} & \textbf{$m^{\mathcal{B}}_1$} & \textbf{$m^{\mathcal{B}}_2$} & \textbf{$m^{\mathcal{B}}_3$}  
 \\ \hline 
$f$ (Hz)&$1000$&\multicolumn{3}{|c|}{$1000$}&\multicolumn{3}{|c|}{$100$}\\ \hline 
Cov&$0.001I_3$&\multicolumn{3}{|c|}{$0.001I_3$}&\multicolumn{3}{|c|}{$0.01I_3$}\\ \hline 
Case $1$&$\checkmark$&$\checkmark$ & $\checkmark$ & $\checkmark$ & $\checkmark$&$\checkmark$ &$\checkmark$\\ \hline
Case $2$&$\checkmark$& $\checkmark$ &$\checkmark$& $\times$&$\times$ & $\checkmark$ & $\times$ \\ \hline 
Case $3$&$\checkmark$& $\times$ &$\times$ & $\checkmark$&$\checkmark$ & $\times$ &$\checkmark$ \\ \hline 
\end{tabular}
\label{table:cases_simulation}
\end{table}
\section{Conclusion}
This work proposes a deterministic framework for attitude estimation using scalar measurements formulated within a linear time-varying (LTV) setting. The observer leverages a linear Kalman filter without linearization, resulting in a robust and simplified design. A closed-form matrix exponential is used in the discrete-time implementation to ensure accurate state transitions under the Zero-Order Hold (ZOH) assumption. The observer's performance is validated through simulations involving partial vector measurements, simulating sensor failures. Future work will focus on extending the approach to a direct design on $\mathrm{SO}(3)$ and on handling disturbances such as biased sensor measurements, as well as improving robustness under diverse sensor configurations. More importantly, we plan to validate the proposed solution in a practical use-case scenario involving failures in some measured directions.


\appendix
\subsection*{A. \textit{Proof of Lemma 1}}
Let us first compute the state transition matrix for \eqref{equation:LTV_state_space}. Let $\mu>0$ and  consider the change of variable $x(t)=T(t)x^{\mathcal{B}}(t)$ where $T(t)=I_3\otimes R(t)$, for any $t\geq0$. Then, by direct differentiation one obtains $\dot x =0$ which implies that $x(t)=x(s)$ for any $0\leq s \leq t$. Therefore, $x^{\mathcal{B}}(t)=T(t)^{\top}T(s)x^{\mathcal{B}}(s)$ which implies that the state transition matrix is given by $\phi(t,s)=T(t)^{\top}T(s)$.
The observability Gramian $W(t, t + \delta)$ for the pair $(A(t), C(t))$ is expressed as:
\begin{multline}\label{equation:proof_oflemma_1_grammian}
W(t, t + \delta) \\=T^\top(t) \left[ \frac{1}{\delta} \int_t^{t+\delta} T(s)C^\top(s)C(s)T^\top(s) \, ds \right] T(t).
\end{multline}
On the other hand, since $T(s) = I_3 \otimes R(s)$ for any $s\geq0$ and in view of \eqref{equation:output_vector}, we obtain $
T(s)C^\top(s)C(s)T^\top(s) = \sum_{i=1}^q (b_i(s) \otimes R(s)a_i(s))(b_i(s) \otimes R(s)a_i(s))^\top$.
Substituting this result back into  \eqref{equation:proof_oflemma_1_grammian}, we obtain \begin{equation}
    W(t, t + \delta) = T^\top(t) \bar{W}(t, t + \delta) T(t),
\end{equation}
where $
\bar{W}(t, t + \delta) =\frac{1}{\delta} \int_t^{t+\delta} \sum_{i=1}^q (b_i(s) \otimes R(s)a_i(s))(b_i(s) \otimes R(s)a_i(s))^\top \, ds.$ In view of \eqref{equation:condition_of_lemma_1}, we have $\bar{W}(t, t + \delta)\geq\mu I_9$.
 Therefore, $W(t,t+\delta)\geq \mu I_{9}$, and thus, the pair $(A(t),C(t))$ is  uniformly observable. This concludes the proof.

\subsection*{B. \textit{Proof of Lemma 2}}
Using the identity $(A\otimes B)(C\otimes D)=(AC)\otimes(BD)$ and, in view of Lemma~\ref{Lemma:general_condition}, one obtains
\begin{align*}
     \sum_{i=1}^{3m}&\big(b_i \otimes Ra_i\big)\big(b_i \otimes Ra_i\big)^\top=  \sum_{i=1}^{3m}(b_ib_i^\top)\otimes (Ra_ia_i^\top R^\top)\\
    &=\sum_{j=1}^{m}\left[(r_j^{\mathcal{I}}(r_j^{\mathcal{I}})^\top)\otimes \sum_{i=1}^3(Ra_{i+3(j-1)}a_{i+3(j-1)}^\top R^\top)\right]\\
    &= \sum_{j=1}^{m}\left[(r_j^{\mathcal{I}}(r_j^{\mathcal{I}})^\top)\otimes I_3\right],
\end{align*}
where we have used the fact that 
$e_1e_1^{\top}+e_2e_2^{\top}+e_3e_3^{\top}=I_3$ to obtain the last equality. Therefore, the condition of Lemma~\ref{Lemma:general_condition} is equivalent to the PE condition given in \eqref{equation:condition_of_lemma_2}, which concludes the proof.

\bibliographystyle{IEEEtran} 
\bibliography{references} 

\begin{thebibliography}{10}
\providecommand{\url}[1]{#1}
\csname url@samestyle\endcsname
\providecommand{\newblock}{\relax}
\providecommand{\bibinfo}[2]{#2}
\providecommand{\BIBentrySTDinterwordspacing}{\spaceskip=0pt\relax}
\providecommand{\BIBentryALTinterwordstretchfactor}{4}
\providecommand{\BIBentryALTinterwordspacing}{\spaceskip=\fontdimen2\font plus
\BIBentryALTinterwordstretchfactor\fontdimen3\font minus \fontdimen4\font\relax}
\providecommand{\BIBforeignlanguage}[2]{{%
\expandafter\ifx\csname l@#1\endcsname\relax
\typeout{** WARNING: IEEEtran.bst: No hyphenation pattern has been}%
\typeout{** loaded for the language `#1'. Using the pattern for}%
\typeout{** the default language instead.}%
\else
\language=\csname l@#1\endcsname
\fi
#2}}
\providecommand{\BIBdecl}{\relax}
\BIBdecl

\bibitem{shuster1981three}
M.~D. Shuster and S.~D. Oh, ``Three-axis attitude determination from vector observations,'' \emph{Journal of Guidance and Control}, vol.~4, no.~1, pp. 70--77, 1981.

\bibitem{markley1988attitude}
F.~L. Markley, ``Attitude determination using vector observations and the singular value decomposition,'' \emph{Journal of the Astronautical Sciences}, vol.~36, no.~3, pp. 245--258, 1988.

\bibitem{Mahony_Hamel_Pflimlin}
R.~Mahony, T.~Hamel, and J.~M. Pflimlin, ``Nonlinear complementary filters on the special orthogonal group,'' \emph{IEEE Transactions on Automatic Control}, vol.~53, no.~5, pp. 1203--1218, 2008.

\bibitem{batista2012sensor}
P.~Batista, C.~Silvestre, and P.~Oliveira, ``Sensor-based globally asymptotically stable filters for attitude estimation: Analysis, design, and performance evaluation,'' \emph{IEEE Transactions on Automatic Control}, vol.~57, no.~8, pp. 2095--2100, 2012.

\bibitem{zlotnik2016nonlinear}
D.~E. Zlotnik and J.~R. Forbes, ``Nonlinear estimator design on the special orthogonal group using vector measurements directly,'' \emph{IEEE Transactions on Automatic Control}, vol.~62, no.~1, pp. 149--160, 2016.

\bibitem{tayebi2006attitude}
A.~Tayebi and S.~McGilvray, ``Attitude stabilization of a {VTOL} quadrotor aircraft,'' \emph{IEEE Transactions on Control Systems Technology}, vol.~14, no.~3, pp. 562--571, 2006.

\bibitem{izadi2014rigid}
M.~Izadi and A.~K. Sanyal, ``Rigid body attitude estimation based on the {L}agrange--d’{A}lembert principle,'' \emph{Automatica}, vol.~50, no.~10, pp. 2570--2577, 2014.

\bibitem{crassidis2007survey}
J.~L. Crassidis, F.~L. Markley, and Y.~Cheng, ``Survey of nonlinear attitude estimation methods,'' \emph{Journal of Guidance, Control, and Dynamics}, vol.~30, no.~1, pp. 12--28, 2007.

\bibitem{barrau2017invariant}
A.~Barrau and S.~Bonnabel, ``The invariant extended {K}alman filter as a stable observer,'' \emph{IEEE Transactions on Automatic Control}, vol.~62, no.~4, pp. 1797--1812, 2017.

\bibitem{barrau2018invariant}
------, ``Invariant {K}alman filtering,'' \emph{Annual Review of Control, Robotics, and Autonomous Systems}, vol.~1, no.~1, pp. 237--257, 2018.

\bibitem{hua2013implementation}
M.-D. Hua, G.~Ducard, T.~Hamel, R.~Mahony, and K.~Rudin, ``Implementation of a nonlinear attitude estimator for aerial robotic vehicles,'' \emph{IEEE Transactions on Control Systems Technology}, vol.~22, no.~1, pp. 201--213, 2013.

\bibitem{berkane2017design}
S.~Berkane and A.~Tayebi, ``On the design of attitude complementary filters on $so(3)$,'' \emph{IEEE Transactions on Automatic Control}, vol.~63, no.~3, pp. 880--887, 2017.

\bibitem{berkane2021nonlinear}
S.~Berkane, A.~Tayebi, and S.~D. Marco, ``A nonlinear navigation observer using imu and generic position information,'' \emph{Automatica}, vol. 127, p. 109513, 2021.

\bibitem{Wang_TAC_2022}
M.~Wang, S.~Berkane, and A.~Tayebi, ``Nonlinear observers design for vision-aided inertial navigation systems,'' \emph{IEEE Transactions on Automatic Control}, vol.~67, no.~4, pp. 1853--1868, 2022.

\bibitem{batista2012ges}
P.~Batista, C.~Silvestre, and P.~Oliveira, ``A {GES} attitude observer with single vector observations,'' \emph{Automatica}, vol.~48, no.~2, pp. 388--395, 2012.

\bibitem{batista2012globally}
------, ``Globally exponentially stable cascade observers for attitude estimation,'' \emph{Control Engineering Practice}, vol.~20, no.~2, pp. 148--155, 2012.

\bibitem{benahmed2024universal}
S.~Benahmed and S.~Berkane, ``A generic observer design for inertial navigation systems using an {LTV} framework,'' \emph{arXiv:2410.03846}, 2025.

\bibitem{Ariante_Ponte_Papa_DelCore}
G.~Ariante, S.~Ponte, U.~Papa, and G.~D. Core, ``Estimation of airspeed, angle of attack, and sideslip for small unmanned aerial vehicles ({UAV}s) using a micro-pitot tube,'' \emph{Electronics}, vol.~10, no.~19, p. 2325, 2021.

\bibitem{de2024pitot}
T.~L. de~Oliveira, P.~van Goor, T.~Hamel, R.~Mahony, and C.~Samson, ``Pitot tube measure-aided air velocity and attitude estimation in {GNSS} denied environment,'' \emph{European Journal of Control}, vol. 2024, p. 101070, 2024.

\bibitem{Hamel_Samson}
T.~Hamel and C.~Samson, ``Position estimation from direction or range measurements,'' \emph{Automatica}, vol.~82, pp. 137--144, 2017.

\bibitem{loria2002uniform}
A.~Lor{\i}a and E.~Panteley, ``Uniform exponential stability of linear time-varying systems: revisited,'' \emph{Systems \& Control Letters}, vol.~47, no.~1, pp. 13--24, 2002.

\bibitem{sarabandi2023solution}
S.~Sarabandi and F.~Thomas, ``Solution methods to the nearest rotation matrix problem in $\mathbb{R}^3$: A comparative survey,'' \emph{Numerical Linear Algebra with Applications}, vol.~30, no.~5, p. e2492, 2023.

\bibitem{trumpf2012analysis}
J.~Trumpf, R.~Mahony, T.~Hamel, and C.~Lageman, ``Analysis of non-linear attitude observers for time-varying reference measurements,'' \emph{IEEE Transactions on Automatic Control}, vol.~57, no.~11, pp. 2789--2800, 2012.

\end{thebibliography}

\end{document}